\documentclass[a4,12pt,fleqn]{article}
\usepackage{epsfig,times,latexsym,enumerate,lscape,float,flafter,amsmath}

\newcommand{\prd}{Phys. Rev. D } 
\newcommand{\mnras}{MNRAS }
\newcommand{\cqg}{Class. Quantum Grav. }
\newcommand{\aaa}{Astron. and Astrophys. }

\begin{document}
\title{Matching of the continuous gravitational wave in an all sky search}
\author{S.K. Sahay\thanks{Present address: Inter University Centre for 
Astronomy and Astrophysics, Post Bag 4, 
Ganeshkhind, Pune-411007, India}\hspace{2.0mm}\thanks{E-mail: ssahay@iucaa.ernet.in}\\ 
Department of Physics, DDU Gorakhpur University, Gorakhpur-273009, India}
\date{}

\maketitle

\begin{abstract}
 
We investigate the matching of continuous gravitational wave (CGW)
signals in an all sky search with reference to Earth based laser 
interferometric detectors. We consider the source location as the 
parameters of the signal manifold and templates corresponding 
to different source locations. It has been found that the matching of
signals from locations in the sky that differ in their co-latitude
 and longitude by $\pi$ radians decreases with source frequency. We have also 
made an analysis with the other parameters affecting the symmetries. We 
observe that it may not be relevant to take care of the symmetries in the 
sky locations for the search of CGW from the output of LIGO-I, GEO600 and TAMA 
detectors.

\end{abstract}

\section{Introduction}
\par The first generation of kilometer-scale gravitational wave (GW) 
laser interferometric detectors 
with sensitivity in the frequency band 10 Hz to few kHz and ultra cryogenic 
bar detectors sensitive at frequencies around 1 kHz will start collecting data 
soon. The TAMA 300 (Tsubona 1995) has 
already done the first large scale data acquisition (Tagoshi et al. 2001), 
while LIGO (Abramovici  et al. 1992) and GEO600 (Danzmann 1995) have recently carried out their 
first science observations. VIRGO (Bradaschia et al. 1991) may become operational in couple of years. 
Also, an eighty meter 
research interferometer ACIGA (McCleland et al. 2000) near Perth, Australia is 
under construction, hoping that it may be possible to extend it to multi-kilometer scale in the future. 

\par  At present, majority of searches are focussed in the detection of {\it 
chirp\/} and burst signals. However interest for the search of continuous 
gravitational wave (CGW) signals from the output of detectors is growing 
(Jaranwoski, et. al 1998, Jaranowski \& Kr\'{o}lak 1999, 2000; Astone et al. 
2002; Brady et. al. 1998, Brady \& Creighton 2000) due to the possibility of 
enhancing the signal-to-noise ratio (S/N) by square root of observation time 
$\sqrt{T_{obs}}$. An optimistic estimate suggests that the Earth based laser 
interferometric detectors may detect such signals with an 
observation time of 1-yr.

\par The strength of the CGW largely depend on the degree of 
long-lived asymmetry in the source. There are several mechanism for 
producing such an asymmetry (Pandharipande et al. 1976;  Bonazzola \&  
Gourgoulhon 1996; Zimmermann \& Szedenits 
1979; Zimmermann  1980). The estimates of the 
asymmetry in neutron stars shows that the amplitude of CGW may be 
$\le 10^{-25}$. Hence, long integration will be required to get the 
signature of signal. But this in turn induces several 
other problems viz. Doppler modulation and non-stationarity of the 
detector noise. Consequently data analysis becomes more harder. However, 
Doppler modulation will provide the information of position of the source in 
the sky. 
 
\par It has been realized that with present computing power, coherent 
all-sky full frequency search in the bandwidth (BW) of the detector of a 
month long data is computationally prohibitive. Some computationally efficient 
alternative approaches have been suggested viz.  
{\it tracking\/} and {\it stacking\/} (Brady and Creighton 2000; 
Schutz 1998). {\it Tracking \/} involves the tracking of lines in the 
time-frequency plane built from the Fourier transform (FT) of one day 
long stretches of data while {\it stacking\/} involves dividing the 
data into day long stretches, searching each stretch for signals, and 
enhancing the detectability by incoherently summing the FT of 
data stretches. Also, accurate modeling of GW form, optimal data 
processing and efficient programming are an integral part 
of all sky-search.  

\par The basic method to analyze the detector output to get the 
signature of GW signals rely on how efficiently one can 
Fourier analyze the data. Fourier analysis of the data has the 
advantage of incorporating the interferometers noise spectral density. 
The problem for the search of CGW largely depend on how accurately one 
can take into account the translatory motion of 
the detector acquired from the motions of the Earth in Solar system barycentre 
(SSB) frame. It has 
been shown (Srivastava and Sahay 2002a,b) that amplitude modulation 
will only redistribute the power of the frequency modulated (FM) signal in 
five frequency bands 
$f \pm 2f_{rot}$, $f$, $f \pm f_{rot}$, where $f$ and $f_{rot}$ are the 
frequencies of the FM signal and the rotational frequency of the Earth 
respectively. Hence it is sufficient to consider only FM signal for the 
analysis of the matching of signals from different locations in the sky . 

\par The study of the matching of
signals from locations in the sky that differ in their co-latitude
 and longitude by $\pi$ radians has been made for 1-d data set (Srivastava 
and Sahay 2002c). They observed symmetries in the sky 
locations. In this paper we report that the observed symmetries are not 
generic. We study this problem for longer observation time and the other  
parameters which affect the symmetries. In the next Section we 
briefly review the FT of FM signal. In Section 3, using the concept of 
fitting factor (FF), we investigate the matching of signals in an all 
sky-search with reference to the Earth based laser 
interferometric detector by considering the source location as the 
parameter of signal manifold and templates corresponding to different 
source locations for longer observation time. We present our conclusions 
in Section 4.

\section{Fourier transform of the frequency modulated continuous gravitational 
wave}
The time dependence of the  phase of a monochromatic CGW signal of frequency 
$f_o$ observed at detector location is given as (Srivastava and Sahay 2002a)
\begin{eqnarray}
\Phi (t) & = &  2\pi f_o \left[ t + {R_{se}\over c} \sin
\theta\cos\phi' + \right.
{R_e\over c}\sin\alpha \{\sin\theta (\sin\beta\cos\epsilon\sin\phi + 
\cos\phi\cos\beta) + \nonumber \\
&& \sin\beta\sin\epsilon\cos\theta\} -  {R_{se}\over c}\sin\theta\cos\phi - 
 {R_e\over c}\sin\alpha\{\sin\theta (\sin\beta_o\cos\epsilon\sin\phi + \nonumber \\ 
&& \cos\phi\cos\beta_o) + \left. \sin\beta_o\sin\epsilon\cos\theta\} \right]\nonumber \\
& = & 2\pi f_o t + {\cal Z}\cos (a\xi_{rot} - \phi ) + 
{\cal N}\cos (\xi_{rot} - \delta ) - {\cal R} - {\cal Q}
\label{eq:phit}
\end{eqnarray}
\noindent where
\begin{equation}
\label{eq:pq}
\left.\begin{array}{lcl}
\vspace{0.2cm}
{\cal P}& = & 2\pi f_o {R_e\over c} \sin\alpha (\cos\beta _o(\sin
\theta \cos\epsilon \sin\phi + \cos\theta \sin\epsilon )
 - \sin\beta _o \sin\theta \cos\phi )\, ,\\
\vspace{0.2cm}
{\cal Q}& = & 2\pi f_o {R_e\over c}\sin\alpha (\sin\beta _o (\sin\theta \cos
\epsilon \sin\phi + \cos\theta \sin\epsilon )
+ \cos\beta _o \sin\theta\cos\phi ) \, ,\\
\vspace{0.2cm}
{\cal N}& = & \sqrt{ {\cal P}^2 + {\cal Q}^2 }\, ,\\
\vspace{0.2cm}
{\cal Z}& = & 2\pi f_o {R_{se}\over c}\sin\theta \, , \\
{\cal R}& = & {\cal Z}\cos\phi\, ,\\
\end{array} \right\}
\end{equation}
\begin{equation}
\label{eq:daydelta}
\left.\begin{array}{lcl}
\vspace{0.2cm}
\delta & = &  \tan^{- 1}\frac{{\cal P}}{{\cal Q}}\, ,\\
\vspace{0.2cm}
\phi' & = & w_{orb}t - \phi\, ,\\
\vspace{0.2cm}
\beta &=& \beta_o + w_{rot} t\, ,\\
\vspace{0.2cm}
\xi_{orb} & = & w_{orb}t\; = \; a\xi_{rot};\quad  a \;= \; w_{orb}/w_{rot}\; \approx \; 1/365.26\, , \\
\xi_{rot} & = & w_{rot}t 
\end{array} \right\}
\end{equation}

\noindent  where $R_{e}$, $R_{se}$, $w_{rot}$ and $w_{orb}$ represent respectively the
Earth's radius, the average distance between the centre of Earth from the 
origin of SSB frame, the rotational and the orbital angular velocity of the 
Earth.  $\epsilon$ and $c$ represent the obliquity of the ecliptic and the velocity of
light. $\alpha$ is the colatitude of the detector. Here $t$ represents the
time in s elapsed from the instant the Sun is at the Vernal Equinox and $\beta_o$
is the local sidereal time at that instant, expressed in radians. $\theta$ and 
$\phi$ denote the celestial colatitude and celestial longitude
of the source. These coordinates are related
to the right ascension, $\bar{\alpha}$ and the declination, $\bar{\delta}$ of
the source via
\begin{equation}
\left.\begin{array} {rcl}
\vspace{0.2cm}
\cos\theta & = & \sin\bar{\delta}\cos\epsilon - \cos\bar{\delta}\sin
\epsilon\sin\bar{\alpha}\\
\vspace{0.2cm}
\sin\theta\cos\phi & = & \cos\bar{\delta}\cos\bar{\alpha} \\
\sin\theta\sin\phi & = & \sin\bar{\delta}\sin\epsilon + \cos\bar{\delta}
\cos\epsilon\sin\bar{\alpha}
\end{array}\right\}
\vspace{0.3cm}
\end{equation}

\noindent The two polarisation states of the signal can be taken as
\begin{equation}
h_+(t) = h_{o_+}\cos [\Phi (t)] 
\label{eq:hpt}
\end{equation}
\begin{equation}
h_\times (t) = h_{o_\times}\sin [\Phi (t)]
\label{eq:hct}
\end{equation}

\noindent $h_{o_+}$, $h_{o_\times}$ are the time independent amplitude of
$h_+(t)$, and  $ h_\times (t)$ respectively.\\ 

\indent For the analysis of the matching of
signals from different locations in the sky, it is sufficient to consider 
either of the polarisation states of FM signal. We consider the $+$ 
polarisation of the signal of unit amplitude. Therefore, the FT for a data 
of $T_{obs}$ observation time may be expressed as (Srivastava and Sahay 2002b)
\begin{eqnarray}
\tilde{h}(f)& =& \int_0^{T_{obs}} \cos[\Phi (t)]e^{-i2\pi ft}dt  \nonumber \\
&& \simeq  \frac{\nu}{2 w_{rot}} \sum_{k  =  - 
\infty}^{k = \infty} \sum_{m = - \infty}^{m =  \infty} e^{ i {\cal A}}{\cal 
B}[ \tilde{{\cal C}} - i\tilde{{\cal D}} ] \; ; \;
\label{eq:hf}
\end{eqnarray}  

\noindent where
\begin{equation}
\left.\begin{array}{lcl}
\vspace{0.2cm}
\nu & = & \frac{f_o - f}{f_{rot}} \\
\vspace{0.2cm}
{\cal A}&  = &{(k + m)\pi\over 2} - {\cal R} - {\cal Q}  \\
\vspace{0.2cm}
{\cal B} & = & {J_k({\cal Z}) J_m({\cal N})\over {\nu^2 - (a k + m)^2}}\\
\vspace{0.2cm}
\tilde{{\cal C}} &=& \sin \nu\xi_o \cos ( a k \xi_o + m\xi_o - k \phi - m \delta )\\
\vspace{0.2cm}
&&  - { a k + m \over \nu}\{\cos\nu\xi_o \sin ( a k \xi_o + m\xi_o - k \phi - m \delta )+ \sin ( k \phi + m \delta )\}\\\
\vspace{0.2cm}
\tilde{{\cal D}} & = & \cos \nu\xi_o \cos ( a k \xi_o + m\xi_o - k \phi - m \delta )   \\
\vspace{0.2cm}
&& + {k a + m \over \nu}\sin \nu \xi_o \sin ( a k \xi_o + m\xi_o - k \phi - m \delta )- \cos ( k \phi + m \delta )\\
\vspace{0.2cm}
\xi_o & = & w_{rot}T_{obs}
\end{array} \right\}
\end{equation}

\noindent J stands for the Bessel function of first kind. The computational strain 
to compute $\tilde{h}(f)$ from equation~(\ref{eq:hf}) can be reduced by $\approx 50\%$ 
by using the symmetrical property of the Bessel functions, given as 
\begin{eqnarray}
\tilde{h}(f)&\simeq & { \nu \over w_{rot}}\left[ {J_o({\cal Z}) J_o
({\cal N}) \over 2\nu^2}\left[ \{ \sin ( {\cal R} + {\cal Q} ) - \sin
({\cal R} + {\cal Q} - \nu\xi_o )\}\; + \right. \right. \nonumber\\ &&
i \left. \{ \cos ( {\cal R} + {\cal Q} ) - \cos ({\cal R} + {\cal Q} - \nu
\xi_o )\} \right]\; + \nonumber \\
&& J_o ({\cal Z})\sum_{m = 1}^{m = \infty} {J_m({\cal N})\over 
\nu^2 - m^2} \left[ ( {\cal Y} {\cal U} -  {\cal X} {\cal V} ) - i ( 
{\cal X} {\cal U} + {\cal Y} {\cal V} ) \right]\; + \nonumber \\ 
&& \left.\sum_{k = 1 }^{k = \infty}\sum_{m = -
\infty}^{m = \infty} e^{ i {\cal A}}{\cal B}\left(
\tilde{{\cal C}} - i\tilde{{\cal D}} \right)\right]\; ;
\label{eq:fm_code}
\end{eqnarray} 

\begin{equation}
\left.\begin{array}{ccl}
{\cal X}& =& \sin ({\cal R}  + {\cal Q} - m \pi/2 )\\
{\cal Y}& =& \cos ({\cal R} +{\cal Q} - m \pi/2 )\\
{\cal U}& =& \sin \nu\xi_o \cos m ( \xi_o - \delta ) - {m\over \nu}\left\{\cos
\nu\xi_o \sin m ( \xi_o - \delta ) - \sin m\delta\right\}\\
{\cal V}& =& \cos \nu\xi_o \cos m ( \xi_o - \delta ) + {m\over \nu}\sin 
\nu\xi_o \sin m ( \xi_o - \delta ) - \cos m\delta\\
\end{array}\right\}
\end{equation}

Equation~(\ref{eq:fm_code}) contains double infinite series of Bessel function. 
However, we know that the value of Bessel function decreases rapidly as its order
exceeds the argument. From equations~(\ref{eq:pq}), it can be shown  
that the order of Bessel function required to compute 
$\tilde{h}(f)$ in the infinite series are
\begin{equation}
k\approx 3133.22 \times 10^3 \sin\theta \left(\frac{f_o}{1 \text{kHz}}\right)\, ,
\end{equation}
\begin{equation}
m\approx 134 \left(\frac{f_o}{1 \text{kHz}}\right).
\end{equation}

\section{Matching of the continuous gravitational wave}
Matched filtering is the most suitable technique for the detection of signals
whose wave form is known. The wave forms are used
to construct a bank of templates, which represent the expected signal wave form
with all possible ranges of its parameters (source location, ellipticity, 
etc.). In an all sky search by match filtering, Srivastava and Sahay 2002c 
observed symmetries under the following transformations
\begin{equation}
\theta_T \longrightarrow  \pi - \theta_T  \qquad 0 \le \theta_T \le \pi\, ,
\label{eq:sym_theta}
\end{equation}
\begin{equation}
\phi_T \longrightarrow \pi - \phi_T \qquad 0 \le \phi_T \le \pi\, ,
\label{eq:sym_phi1}
\end{equation}
\begin{equation}
\phi_T \longrightarrow 3\pi - \phi_T \qquad \pi \le \phi_T \le 2\pi . 
\label{eq:sym_phi2}
\end{equation}

\par The above observations are made by studying few cases for 1-d data set. 
Hence, it 
will be important to understand and check the generic nature of the matching 
of signals under the above transformations. To study this problem we use 
the formula for FF (Apostolatos 1995) which quantitatively describes the 
closeness of two signals, given as
\begin{eqnarray} 
\text{FF} & = & \frac{\langle h(f)|
h_T(f;\theta_T , \phi_T)\rangle}{\sqrt{\langle h_T(f;\theta_T , \phi_T )|h_T(f;
\theta_T , \phi_T )\rangle\langle h(f)|h(f)\rangle}}
\label{eq:ff1}
\end{eqnarray}

\noindent where $h(f)$ and $h_T(f; \theta_T , \phi_T)$ represent respectively the
FTs of the actual signal wave form and the templates. The inner product of two 
waveform $h_1$ and $h_2$ is defined as
\begin{eqnarray}
\langle h_1|h_2\rangle & =& 2\int_0^\infty \frac{\tilde{h}_1^*(f)\tilde{h}_2(f)
+ \tilde{h}_1(f)\tilde{h}_2^*(f)}{S_n(f)}df \nonumber \\
 & = &
4\int_0^\infty \frac{\tilde{h}_1^*(f)\tilde{h}_2(f)}{S_n(f)}df
\label{eq:ip}
\end{eqnarray}

\noindent where $^*$ denotes complex conjugation, $\tilde{}$ denotes the FT of the quantity 
underneath 
$(\tilde{a}(f) = \int_\infty^\infty a(t)\text{exp}(- 2 \pi i f t)dt) $ and $S_n(f)$ is the 
spectral density of the detector noise. In our analysis we assume the 
noise to be stationary and Gaussian. It is remarked that to compute the 
inner product of the signals one would require the 
BW of the Doppler modulated signal. From the Doppler shift  
(Srivastava and Sahay 2002a), the BW of Doppler modulated signal may be 
given as  
\begin{equation}
\text{BW} \approx (1.99115 \times 10^{-4}\sin\theta + 3.09672 
\times 10^{-6}) f_o.
\label{eq:bandwidth}
\end{equation}

\subsection{Celestial colatitude}
\begin{figure}
\centering
\epsfig{file=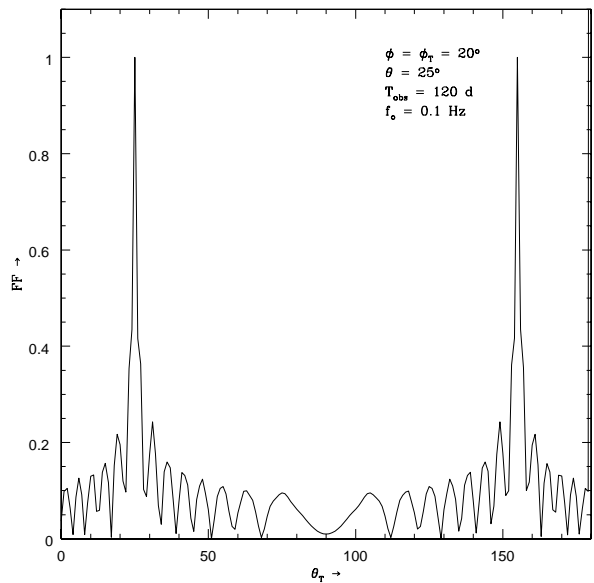,height=8.40cm}
\caption{Variation of FF with $\theta_T$.}
\label{fig:theta_var}
\end{figure}

\begin{figure}
\centering
\epsfig{file=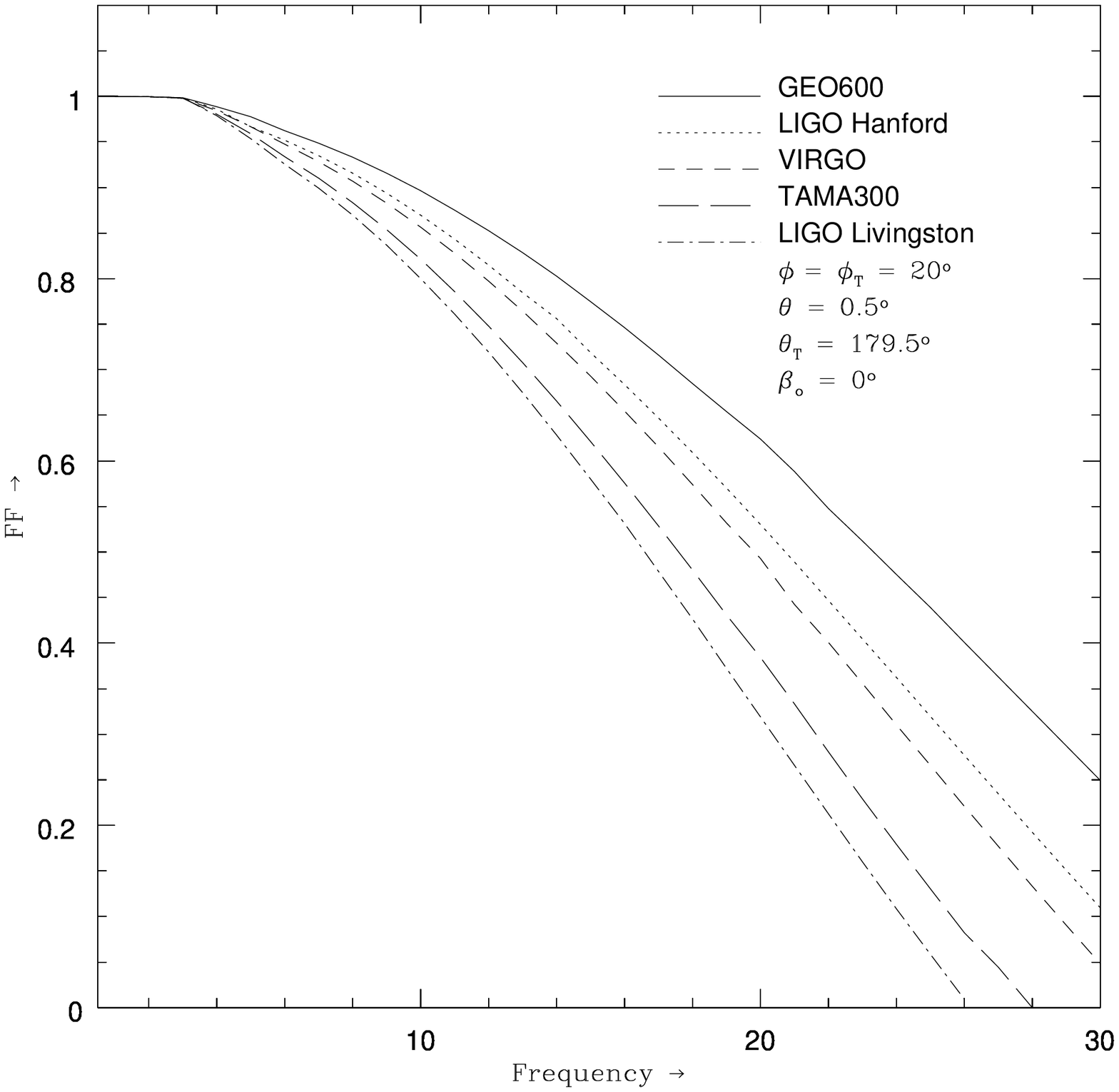,height=8.40cm}
\caption{Fall of FF with $f_o$.}
\label{fig:freq_beta0}
\end{figure}

\begin{figure}
\centering
\epsfig{file=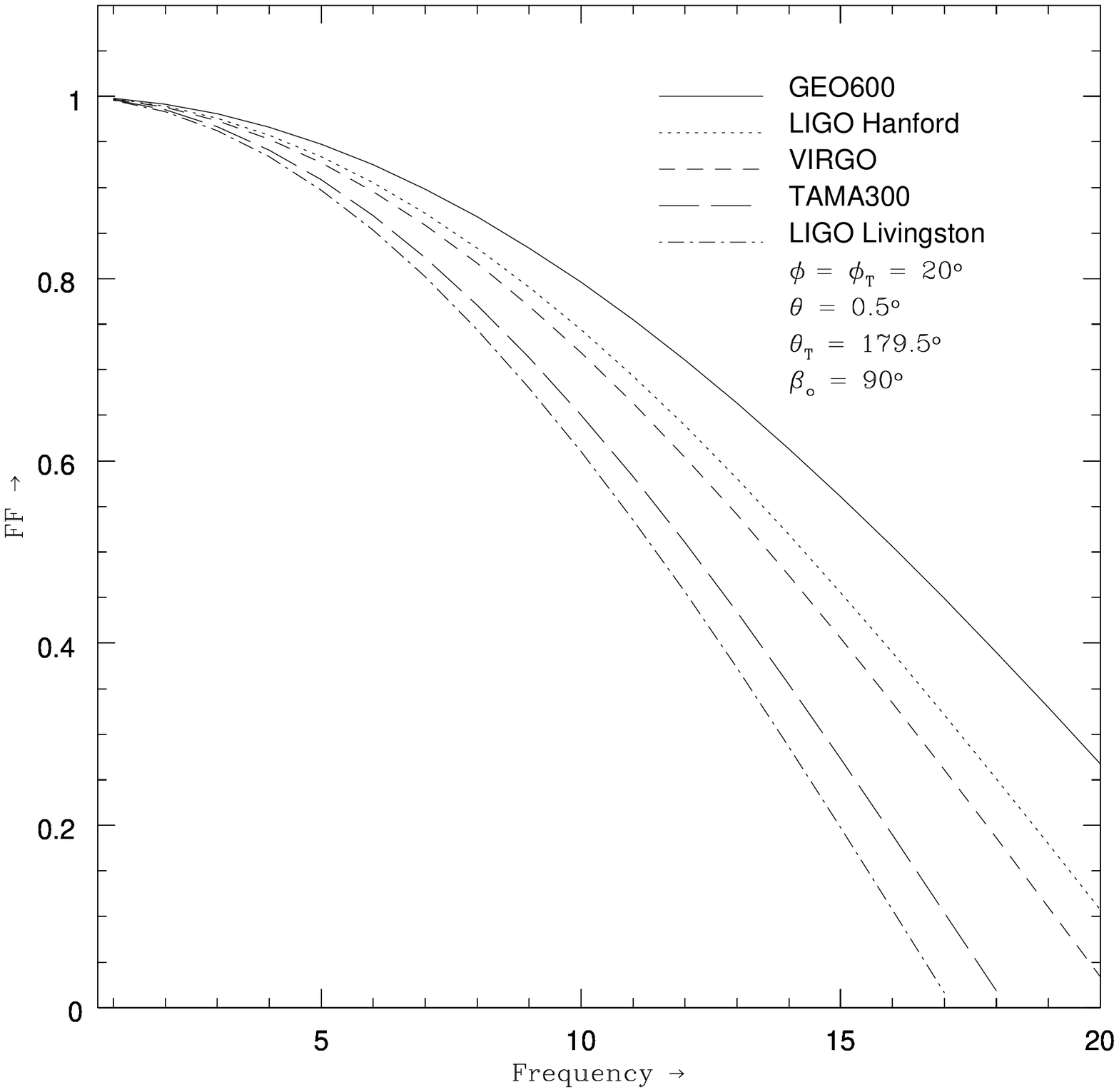,height=8.40cm}
\caption{Fall of FF with $f_o$.}
\label{fig:freq_beta90}
\end{figure}

Let us consider that GEO600 detector (the position and orientation of the 
detectors can be found in Jaranowski et al. 1998) receive a CGW signal of 
frequency $f_o=0.1$ Hz (unreasonably low frequency has been chosen for 
illustrative purposes limited by available computational power) from a source 
located at $(\theta , \phi) = (25^o,20^o)$. In order to evaluate the matching 
of the signals in colatitude, we 
first maximize FF over $\phi$ by choosing $\phi = \phi_T = 20^o$. Now, we 
wish to check the symmetries in colatitude represented by 
equation~(\ref{eq:sym_theta}) for the data set $T_{obs} = 120$ d. For 
the purpose we maximize the FF over $\theta$  by varying $\theta_T$ in 
discrete steps over entire range i.e. $0^o$ to $180^o$. For the present 
case it is sufficient to take the ranges of $k$ and 
$m$ as $1$ to $345$ and $- 3$ to $3$ respectively and $\text{BW} =  20.1954 
\times 10^{-6} \text{Hz}$. The results so obtained are shown in 
Figure~(\ref{fig:theta_var}). To 
establish the observed symmetries, we similarly compute the FF for $T_{obs} 
= 1,2,3....365$ d and observe that the matching of signals remains almost 
same. However, due to obliquity of the ecliptic, the variation in the matching 
of signals 
will depend on source frequency, colatitude and detector position 
and orientation. We check the dependence of the FF on this parameters. The 
result so obtained for different Earth based laser interferometric detectors 
are shown in the Tables~(\ref{table:theta_var_ff}), \& ~(\ref{table:coeff_theta}) and Figures~(\ref{fig:freq_beta0}) \& ~(\ref{fig:freq_beta90}).

\begin{table}
\centering
\begin{tabular}{|c|c|c|c|cc|c|c|c|c|}
\cline{1-4} \cline{7-10}

$\theta^o$ & $\theta^o_T$ & FF &  FF &\hspace{0.50in} &&$\theta^o$ & $\theta^o_T$ &  FF &FF\\
&&($\beta_o = 0^o$)&($\beta_o = 90^o$)& &&&&($\beta_o = 0^o$)&($\beta_o = 90^o$)\\
\cline{1-4} \cline{7-10}

0.5& 179.5 & 0.9999 &0.9970&&& 45& 135 & 0.9992&0.9939\\
1& 179 & 0.9999 &0.9970&&&  50& 130 &0.9993 &0.9933\\
5& 175 & 0.9992 &0.9970&&&  55& 125 &0.9995 &0.9985\\
10& 170 & 0.9985 &0.9968&&& 60& 120 &0.9996 &0.9988\\
15& 165 & 0.9986 &0.9966&&& 65& 115 & 0.9997&0.9992\\
20& 160 & 0.9987 &0.9963&&& 70& 110 & 0.9998&0.9994\\
25& 155 & 0.9987 &0.9959&&& 75& 105 &0.9998 &0.9997\\
30& 150 & 0.9988 &0.9954&&& 80& 100 &0.9999 &0.9998\\
35& 145 & 0.9990 &0.9949&&& 85& 95 &0.9999 &0.9999\\
40& 140 & 0.9991 &0.9944&&& 89& 91 &0.9999 & 0.9999\\

\cline{1-4} \cline{7-10}
\end{tabular}
\vspace{0.50cm}
\caption{Matching of the signals of frequency 1 Hz under the transformation represented by equation~(\ref{eq:sym_theta}) for GEO600 detector.}
\label{table:theta_var_ff}
\end{table}

\begin{table}
\centering
\begin{tabular}{|l|c|c|c|c|}
\hline
Detector & $A_o$ & $A_1$  & $A_2$  &$ A_3$ \\
 &$\times 10^{- 3}$&$\times 10^{- 5}$&$\times 10^{- 5}$&$\times 10^{- 7}$\\
\hline
$\text{GEO600}_{\left(\beta_o = 0^o\right)}$&1000.02 & 124.524 & 129.030 &137.774\\
$\text{GEO600}_{\left(\beta_o = 90^o\right)}$&997.796 &175.426 &248.072 & 282.478\\
$\text{LIGO Hanford}_{\left(\beta_o = 0^o\right)}$&998.450 & 225.359 & 172.731 &221.187\\
$\text{LIGO Hanford}_{\left(\beta_o = 90^o\right)}$&996.668 &266.811 &324.073 &440.729\\
$\text{VIRGO}_{\left(\beta_o = 0^o\right)}$&998.548 &243.480 &191.503 &259.026\\
$\text{VIRGO}_{\left(\beta_o = 90^o\right)}$&996.064 &316.288 &362.630 &529.531\\
$\text{TAMMA300}_{\left(\beta_o = 0^o\right)}$&997.914 &342.746 &249.527 &390.874\\
$\text{TAMMA300}_{\left(\beta_o = 90^o\right)}$&995.744 &371.193 &458.856 &756.536\\
$\text{LIGO Livingston}_{\left(\beta_o = 0^o\right)}$&998.815 &396.783 &285.368 &478.932\\
$\text{LIGO Livingston}_{\left(\beta_o = 90^o\right)}$&995.622 &399.087 &516.416 &904.211\\
\hline
\end{tabular}
\caption{Coefficients of the fall of FF with $f_o$ under the transformation represented by equation~(\ref{eq:sym_theta}) for $\beta_o = 0^o$ and $90^o$.}
\label{table:coeff_theta}
\end{table}

\noindent The analysis of the matching of signals in $\theta$ space shows that
\begin{enumerate}[(i)]
\item for fixed $f_o$, the FF
\begin{enumerate}
\item is independent of $T_{obs}$ and $\phi$.
\item  does not vary significantly with 
the variation of source location, detector position and orientation 
[Tables~(\ref{table:theta_var_ff}), and~(\ref{table:coeff_theta})].
\end{enumerate}
\item the FF falls with the source frequency [Figures~(\ref{fig:freq_beta0}) 
and~(\ref{fig:freq_beta90})]. From the figure we find that 
 it may not be relevant to take care of the symmetries in the sky locations 
for the search of CGW from the output of LIGO-I, GEO600 and TAMA detectors 
whose lower cut off frequency is 40/75 Hz (Owen and Sathyaprakash 1999).
\item the approximate fall of FF based on the 
figures~(\ref{fig:freq_beta0})and~(\ref{fig:freq_beta90}) may be given as
\end{enumerate}
\begin{equation}
\text{FF} = A_o + A_1 \left(\frac{f_o}{\text{Hz}}\right) - A_2 
\left(\frac{f_o}{\text{Hz}}\right)^2  + A_3 \left(\frac{f_o}{\text{Hz}}\right)^3
\label{eq:ff_coeff1}
\end{equation}
\noindent where $A_o$, $A_1$, $A_2$, $A_3$ are constants given in Table~(\ref{table:coeff_theta}).
\subsection{Celestial longitude}
\begin{figure}
\centering
\epsfig{file=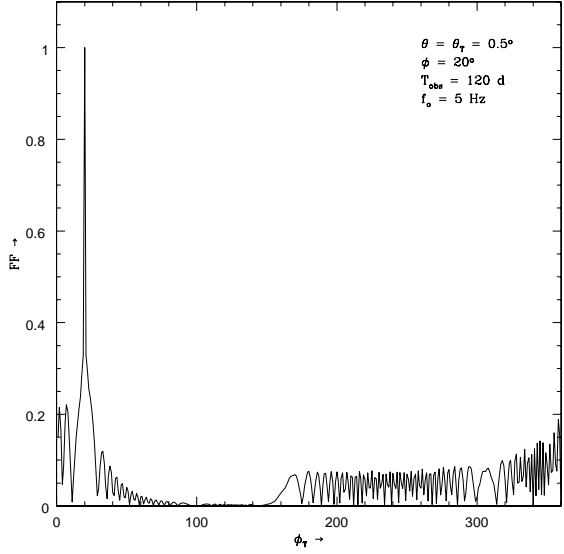,height=8.0cm}
\caption{Variation of FF with $\phi_T$.}
\label{fig:phi_var1}
\end{figure}

\begin{figure}
\centering
\epsfig{file=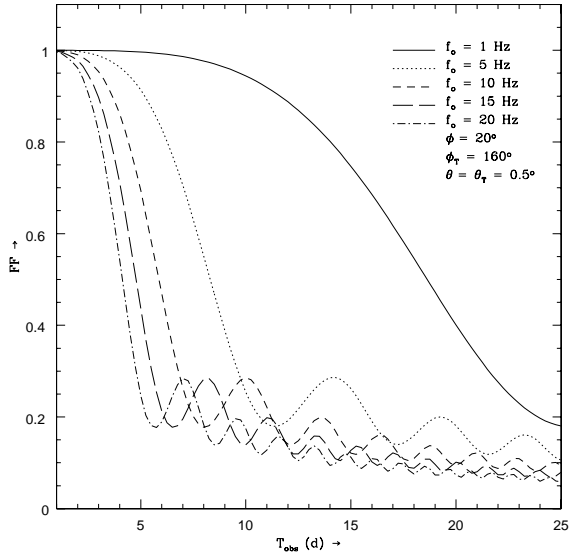,height=8.0cm}
\caption{Fall of FF with $T_{obs}$ for different $f_o$.}
\label{fig:freq_var1}
\end{figure}

\begin{figure}
\centering
\epsfig{file=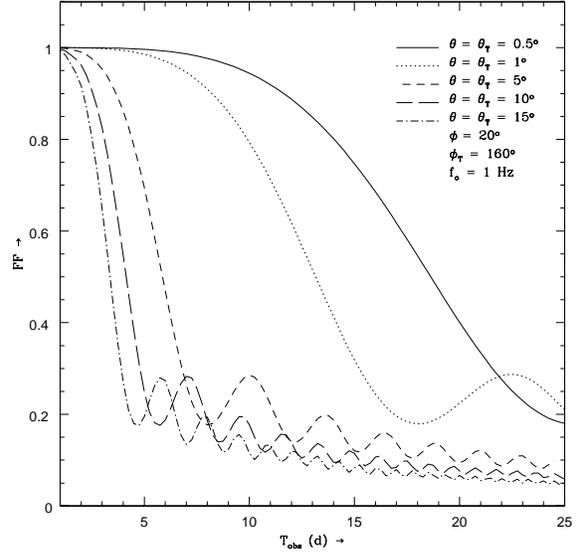,height=8.0cm}
\caption{Fall of FF with $T_{obs}$ for different $\theta$.}
\label{fig:theta_var1}
\end{figure}

\begin{figure}
\centering
\epsfig{file=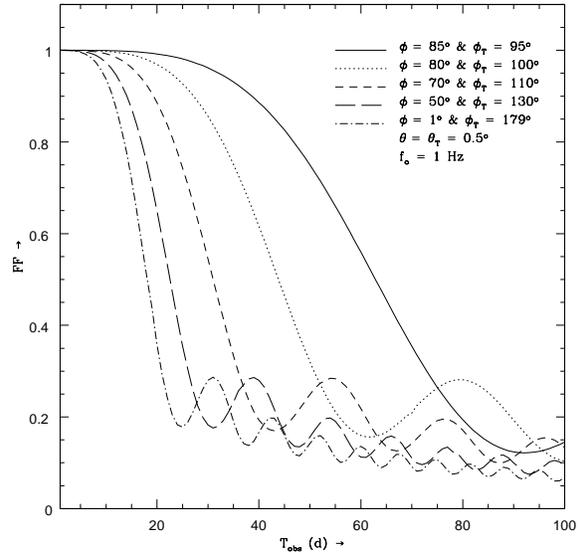,height=8.0cm}
\caption{Fall of FF with $T_{obs}$ for different $\phi$ and $\phi_T$.}
\label{fig:phi_var3}
\end{figure}

The Doppler shift due to the motions of the Earth is mainly depend on the  
colatitude and source frequency   
and have very less dependence on the longitude. Consequently, grid 
spacing of the templates for matched filtering for an all sky search 
will insignificantly depend on longitude (Brady \& Creighton 2000). 
Keeping this in view we similarly check the matching of signals under the 
transformation given by the equation~(\ref{eq:sym_phi1}). We 
 chosen the LIGO detector located at Livingston, selected a data set 
$T_{obs} = 120 $ d, $(\theta , \phi) = (0.5^o , 40^o)$, $f_o = 5$ Hz. In 
order to compute the FF, we first maximize equation~(\ref{eq:ff1}) over 
$\theta$ by selecting $\theta = \theta_T = 0.5^o$, followed by  
maximization over $\phi$ in discrete steps over its entire range, 
$0^o \le \phi \le 360^o$. The result so obtained is shown in 
Figure~(\ref{fig:phi_var1}). We also check the mismatch of the signals 
for different $\theta , \phi$ and $f_o$ by computing the 
FF for the data set of $T_{obs} = 1,2,....25/100$ d. The results so obtained 
are shown in Figures~(\ref{fig:freq_var1}),~(\ref{fig:theta_var1}), and~(\ref{fig:phi_var3}) respectively. Almost same behavior has been observe  
for the transformation represented by the equation~(\ref{eq:sym_phi2}).

\par From the figures we note that the matching of the signals in longitude 
decreases with $T_{obs}, f_o, \theta$ and $\phi$. However, behavior of the 
matching of signals are similar in nature and  may be represented by 
\begin{equation}
\text{FF} = B_o - B_1 T_{obs} + B_2 T_{obs}^2 - B_3 T_{obs}^3 + B_4 T_{obs}^4; 
\qquad T_{obs}= 1,2,....25/100 \;\text{d}
\label{eq:ff_coeff2}
\end{equation}
\par Where, $B_o$, $B_1$, $B_2$, $B_3$, $B_4$ are the constants given in 
Table~(\ref{table:coefficients}). 
The equations~(\ref{eq:ff_coeff1}) and~(\ref{eq:ff_coeff2}) does not 
represents the oscillatory part of the figures. This oscillatory 
behavior is more or less typical in waveform that match well or bad depending 
on their parameters and is low enough in comparison to the threshold  
of the detection of GW that one may choose.
\begin{table}
\centering
\begin{tabular}{|c|c|c|c|c|c|c|c|c|c|}
\hline
$f_o$ &$\theta^o$ &$\theta^o_T$ & $\phi^o$ &  $\phi^o_T$ & $B_o$ & $B_1$  & $B_2$  &$ B_3$  &$ B_4$ \\
(Hz)&&&&&$\times 10^{- 3}$&$\times 10^{- 5}$&$\times 10^{- 5}$&$\times 10^{- 6}$&$\times 10^{- 7}$\\
\hline
20 &&&&&1032.39 & 13167.8 & 14036.2 & 56606.5 & 55088.4\\
 
15 &&&&&1030.85 & 11101.8 & 10398.0 & 36450.9 & 30710.3\\

10 &0.5&0.5&&&1028.79 & 8750.18 & 6813.74 & 19603.4 & 13481.1\\

5 &&&&&1025.73 & 5842.30 & 3302.21 & 6795.19 & 3310.22\\

&&&20/200&160/340&1021.20 & 2348.47 & 616.661 & 579.752 & 126.575\\

&1&1&&&1023.31 & 3527.06 & 1287.73 & 1692.54 & 522.566\\

&5&5&&&1028.85 & 8758.50 & 6812.31 & 19567.0 & 13437.3\\

&10&10&&&1032.67 & 13220.3 & 14031.8 & 56298.5 & 54579.0\\

&15&15&&&1034.40 & 16673.7 & 21157.0 & 103289.0 & 122385.0\\

1&&&1/181&179/359&1031.20 & 2896.18 &722.052 & 671.331 & 149.659\\

&&&50/230&130/310&1028.29 & 2216.41 & 452.688 & 341.054 & 61.1078\\

&0.5&0.5&70/250&110/290&1024.32 & 1473.47 & 225.270 & 125.891 & 16.4451\\

&&&80/260&100/280&1019.87 & 906.855 & 101.752 & 41.5159  & 3.84672\\

&&&85/265&95/275&1012.58&449.523&38.5174&11.9112&0.774374\\

\hline
\end{tabular}
\caption{Coefficients of the fall of FF for LIGO Livingston detector under the transformation represented by equations~(\ref{eq:sym_phi1}) and~(\ref{eq:sym_phi2}) for different $\theta$ and $f_o$.}
\label{table:coefficients}
\end{table}

\twocolumn

\onecolumn
\section{Conclusion}
In view of blind all sky search for CGW, we have studied the matching of  
signals from different source locations assuming the noise to be stationary 
and Gaussian. For fixed $f_o$, we observed that 
the matching of signals from locations in the sky that differ in their 
colatitude by $\pi$ radians is independent of 
$T_{obs}$, and $\phi$. However, it falls with $f_o$. 
But in longitude the matching of signals falls 
with $T_{obs}$, $\theta$, $\phi$, $f_o$. We believe that the matching of 
signals will increase for the real data. This is due to the resolution 
provided by the Fast 
Fourier transform (FFT). However, it may not be relevant to account 
this symmetries for the search of CGW from the output of LIGO-I, 
GEO600 and TAMA detectors.
\par This analysis will be more relevant if one performs hierarchical search 
(Mohanty \& Dhurandhar 1996; Mohanty 1998). This search is 
basically a two step search, in first step the detection threshold is 
kept low and in the second step a higher threshold is used. The higher 
threshold is used for those templates which exceed the first step threshold. 
The study of the matching of signals in different source locations will 
be also relevant for Laser interferometer space antenna (LISA) (Hough J., 
1995). The work has been initiated and may be useful for the data analysis of 
CGW. 

\section*{Acknowledgment}
I am thankful to IUCAA for providing hospitality where major part of the work 
was carried out. I am thankful to Prof. S.V. Dhurandhar, IUCAA, Pune for 
his useful discussions and Mr. A. Sengupta, IUCAA, Pune for reading the 
paper.

\end{document}